# Leaky Modes of Solid Dielectric Spheres

Masud Mansuripur[†], Miroslav Kolesik[†], and Per Jakobsen[‡]

[†]College of Optical Sciences, The University of Arizona, Tucson
[‡]Department of Mathematics and Statistics, University of Tromsø, Norway

**Abstract**. In the absence of external excitation, light trapped within a dielectric medium generally decays by leaking out, and also by getting absorbed within the medium. We analyze the leaky modes of solid dielectric spheres by examining solutions of Maxwell's equations for simple homogeneous, isotropic, linearly dispersive media that admit complex-valued oscillation frequencies. We show that, under appropriate circumstances, these leaky modes constitute a complete set into which an initial electromagnetic field distribution inside a dielectric sphere can be expanded. We provide the outline of a completeness proof, and also present results of numerical calculations that illustrate the close relationship between the leaky modes and the resonances of solid dielectric spherical cavities.

**1. Introduction**. A well-polished, solid, smooth, homogeneous, and transparent glass sphere is a good example of material bodies which, when continually illuminated, admit and accommodate some of the incident light, eventually reaching a steady state where the rate of the incoming light equals that of the outgoing. By properly adjusting the frequency of the incident light, one can excite resonances, thus arriving at conditions under which the optical intensity inside the dielectric host exceeds, often by a large factor, that of the incident light beam [1,2]. If now the incident beam is suddenly terminated, the light trapped within the host medium begins to leak out, and, eventually, that portion of the electromagnetic (EM) energy which is not absorbed by the host, returns to the surrounding environment.

The so-called leaky modes of a dielectric body are characterized by a unique set of complex-valued frequencies $\omega_q = \omega'_q + i\omega''_q$, where the index $q$ is used here to enumerate the modes [3-7]. The imaginary part $\omega''_q$ of each such frequency signifies the decay rate of the leaky mode, and (aside from a numerical coefficient) the corresponding quality-factor is given by $Q = |\omega'_q/\omega''_q|$.

The leaky modes of dielectric waveguides and cavities have been studied for many years, and a considerable volume of results pertaining to these modes exists in the literature. In addition to their applications in computational photonics and electromagnetics [8,9], such states of the EM field also pose questions of fundamental interest. Specifically, the problem of completeness and the general mathematical properties of these so-called "quasi-normal modes" have been broadly investigated. Of particular relevance to the present paper are the results reported in [5], which show that the leaky modes of a dielectric cavity can serve as a basis to represent arbitrary functions but only inside the cavity. In [6] it was shown that the set of leaky modes remains complete in the aforementioned sense even when the host medium exhibits losses as well as some chromatic dispersion limited to finite frequencies. Considering that the inclusion of chromatic dispersion and optical loss complicates the problem considerably, most of the pertinent mathematical analysis to date has been limited to one-dimensional systems, with rigorous results usually associated with cases in which chromatic dispersion and/or optical loss have been absent [10,11].

A specific application of leaky modes is the evaluation of the Purcell spontaneous emission enhancement factor when a dipole oscillator is coupled to a nearby cavity or a plasmonic resonator. Recent publications [12,13] have shown that the Purcell factor can be estimated from one (or a few) leaky modes, thus showcasing the need for delving into its detailed derivation, which exploits the expansion of Green's tensor in terms of the leaky modes. The completeness of the leaky modes is *assumed* in the aforementioned papers, and references are given to the published literature where completeness has been discussed. In the present paper we emphasize that completeness should not be assumed but must be proven, that the expansion of an initial field distribution into a superposition of leaky modes is not trivial but involves subtleties due to the unusual behavior of the leaky modes in the vicinity of the singularities of the refractive index, and that the expansion coefficients obtained without proper accounting for such singularities could be wrong, resulting in a non-convergent series expansion. We will see how the leaky modes accumulate near one of the singular points of the refractive index. Consequently, if and when the excitation frequency happens to be close to the pole(s) of the refractive index, the assumption that one (or at most a few) leaky modes are sufficient to expand a given field distribution would become questionable.

One of the goals of the present paper is to generalize the previous results in several ways. In particular, we study dielectric spheres with chromatic dispersion and loss properties in the framework of a Lorentz oscillator model, which also properly accounts for the behavior at high frequencies. This is an issue of fundamental importance, because high-frequency asymptotics in fact determine the conditions under which a set of leaky modes can be considered complete. As the frequency increases, the refractive index approaches unity and, for all practical purposes, high-frequency propagating waves cease to experience the presence of the cavity. While this is an important feature that was not part of the analysis in [6], one could reasonably argue that, in the absence of confinement for high-frequency EM waves, leaky modes could only provide approximations but not true resonant-mode expansions for arbitrary functions. Yet another difficulty arising from the dispersive properties of the medium is that the Lorentz oscillator model of the refractive index introduces branch-cuts into the analytic structure of the leaky-mode expansion. While the role of this singularity has been related to the over-completeness of the resonant states [14], the existence of branch-cuts can potentially invalidate arguments supporting completeness. We address these questions rigorously, and demonstrate the convergence of the leaky-mode expansion inside the cavity when a realistic high-frequency wave behavior is properly taken into consideration. In doing so, we also show that the leaky-mode expansion can be constructed in a way that eliminates the potential problem caused by the branch-cuts associated with the refractive index.

The present paper contributes to the mathematics of open systems by providing an alternative completeness proof for leaky modes of solid, homogeneous, isotropic, dispersive dielectric spheres. We put forward a new approach to the completeness analysis that might find applications elsewhere in mathematical physics as well. The main tools of the trade in the existing literature on quasi-normal and resonant modes are invariably Green's functions. In contrast to such conventional approaches, we present a method that relies solely on the analytic properties of the scattering states, thus avoiding any reliance on Green's functions. Ours is a straightforward approach that simplifies the analysis of the leaky-mode expansion in comparison to conventional methods. Also possible are similar proofs of completeness for the leaky modes of parallel-plate dielectric slabs and infinitely-long dielectric cylinders, which we have recently reported in a conference proceedings paper [15].

An important question with regard to the completeness issue is the space of functions that can be expressed as a superposition of leaky modes. Interestingly, this is actually rarely addressed in the context of optical cavities. Note that our analysis is concerned with inherently lossy systems; in other words, not only is the system under investigation open, but also its EM energy content can be dissipated throughout the host medium. Thus, the problem being non-Hermitian, one cannot rely on the completeness of the scattering states as a point of departure when attempting to prove that the set of leaky modes forms a basis for expansion. To address this issue, we provide a constructive specification of the function space that is spanned by the leaky modes. It is specified as the space of EM fields excited within the optical cavity under external illumination. Assuming the external excitation has reached a steady state (after a sufficiently long time) when it is terminated, we proceed to show that the subsequent evolution of the EM field left inside the cavity can be represented by a convergent superposition of the leaky modes.

Our approach provides a direct link between the various ways in which resonant modes can be detected and studied. In particular, the correspondence between resonance conditions, line shapes, and $Q$-factors of a spherical cavity can be readily explored. We present numerical results to illustrate certain general properties of spherical cavities. Last but not least, our results provide insight into features of resonant modes that are of practical interest. By comparing the properties of idealized spherical cavities with those made of realistic (i.e., lossy and dispersive) materials, we will show the way in which the losses inherent to the host material impose limits on the achievable $Q$-factors of solid dielectric spheres.

In the following sections, we analyze the EM structure of the leaky modes of solid dielectric spheres, and examine the conditions under which certain initial field distributions can be decomposed into a superposition of leaky modes. We also present numerical results where the resonance conditions, line shapes, and quality-factors of a spherical cavity are computed; the correspondence between these and the leaky-mode frequencies is subsequently explored.

We begin by describing in Sec. 2 the dispersive properties of linear, isotropic, homogeneous dielectric media whose electric permittivity and magnetic permeability each follow a single Lorentz oscillator model. Then, in Sec. 3, after a summary presentation of vector spherical harmonics, we demonstrate the completeness of the leaky modes of solid dielectric spheres for a special class of initial distributions residing within the spherical cavity. Numerical results showing the connection between the resonances of a dielectric sphere (when illuminated by a tunable source) and the corresponding leaky mode frequencies are presented in Sec. 4. Section 5 provides a summary of the main results of the paper followed by a few concluding remarks.

**2. Refractive index model for a dispersive dielectric**. The simplest dispersive dielectric is a medium whose electric and magnetic dipoles behave as independent Lorentz oscillators, each having their own resonance frequency $\omega_r$, plasma frequency $\omega_p$, and damping coefficient $\gamma$ [16,17]. The electric and magnetic susceptibilities of the material will then be given by

$$\chi_e(\omega) = \frac{\omega_{pe}^2}{\omega_{re}^2 - \omega^2 - \mathrm{i}\gamma_e \omega}, \tag{1a}$$

$$\chi_m(\omega) = \frac{\omega_{pm}^2}{\omega_{rm}^2 - \omega^2 - \mathrm{i}\gamma_m \omega}. \tag{1b}$$

The corresponding refractive index, which is also a function of the frequency $\omega$, will then be

$$n(\omega) = \sqrt{\mu\varepsilon} = \sqrt{(1+\chi_m)(1+\chi_e)} = \sqrt{1 + \frac{\omega_{pm}^2}{\omega_{rm}^2 - \omega^2 - \mathrm{i}\gamma_m \omega}} \times \sqrt{1 + \frac{\omega_{pe}^2}{\omega_{re}^2 - \omega^2 - \mathrm{i}\gamma_e \omega}}$$

$$= \sqrt{\frac{(\omega - \Omega_{1m})(\omega - \Omega_{2m})}{(\omega - \Omega_{3m})(\omega - \Omega_{4m})}} \times \sqrt{\frac{(\omega - \Omega_{1e})(\omega - \Omega_{2e})}{(\omega - \Omega_{3e})(\omega - \Omega_{4e})}}, \tag{2a}$$

where

$$\Omega_{1,2} = \pm\sqrt{\omega_r^2 + \omega_p^2 - \tfrac{1}{4}\gamma^2} - \tfrac{1}{2}\mathrm{i}\gamma, \tag{2b}$$

$$\Omega_{3,4} = \pm\sqrt{\omega_r^2 - \tfrac{1}{4}\gamma^2} - \tfrac{1}{2}i\gamma. \tag{2c}$$

Assuming that $\gamma \ll \omega_r$, the poles and zeros of $\mu(\omega)$ and $\varepsilon(\omega)$ will be located in the lower-half of the complex $\omega$-plane, as shown in Fig.1. The dashed line-segments in the figure represent branch-cuts that are needed to uniquely specify each square-root function appearing on the right-hand side of Eq.(2a). For the sake of simplicity, one might further assume that the branch-cuts of $\sqrt{\mu}$ and those of $\sqrt{\varepsilon}$ do *not* overlap, although, strictly speaking, this restriction is not necessary. Whenever $\omega$ crosses (i.e., moves from immediately above to immediately below) one of these four branch-cuts, the refractive index $n(\omega)$ is multiplied by $-1$. Note also that, in the limit when $|\omega| \to \infty$ (along any straight line originating at $\omega = 0$), the complex entities $\mu(\omega)$, $\varepsilon(\omega)$, and the refractive index $n(\omega)$ will all approach 1.0, while $1 - n^2(\omega)$ approaches $(\omega_{pm}^2 + \omega_{pe}^2)/\omega^2$.

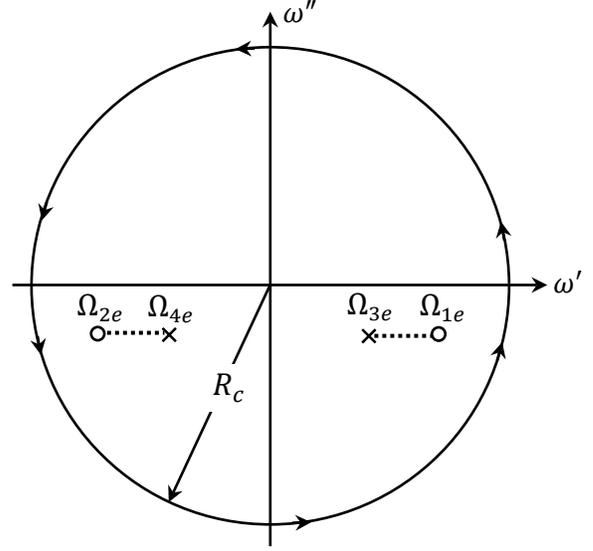

**Fig.1**. Locations in the $\omega$-plane of the poles and zeros of $\varepsilon(\omega)$, whose square root contributes to the refractive index $n(\omega)$ in accordance with Eq.(2). A similar set of poles and zeros, albeit at different locations in the $\omega$-plane, represents $\mu(\omega)$. The dashed lines connecting pairs of adjacent poles and zeros constitute branch-cuts for the function $n(\omega)$. In accordance with the Cauchy-Goursat theorem [18], the integral of a meromorphic function, such as $f(\omega)$, over a circle of radius $R_c$ is $2\pi i$ times the sum of the residues of the function at the poles of $f(\omega)$ that reside within the circle.

**3. Leaky modes of a solid dielectric sphere**. The vector spherical harmonics of the EM field within a homogeneous, isotropic, linear medium having permeability $\mu_0\mu(\omega)$ and permittivity $\varepsilon_0\varepsilon(\omega)$ are found by solving Maxwell's equations in spherical coordinates [16,17]. The electric and magnetic field profiles for Transverse Electric (TE) and Transverse Magnetic (TM) modes of the EM field are found to be

$m = 0$ TE mode ($E_r = 0$):
$$\boldsymbol{E}(\boldsymbol{r},t) = \frac{E_0 J_{\ell+\frac{1}{2}}(kr)}{\sqrt{kr}} P_\ell^1(\cos\theta) \exp(-i\omega t)\widehat{\boldsymbol{\varphi}}. \tag{3}$$

$$\boldsymbol{H}(\boldsymbol{r},t) = \frac{E_0}{\mu_0\mu(\omega)r\omega}\left\{\frac{J_{\ell+\frac{1}{2}}(kr)}{i\sqrt{kr}}[\cot\theta\, P_\ell^1(\cos\theta) - \sin\theta\, \dot{P}_\ell^1(\cos\theta)]\widehat{\boldsymbol{r}}\right.$$
$$\left. - \frac{kr\dot{J}_{\ell+\frac{1}{2}}(kr) + \tfrac{1}{2}J_{\ell+\frac{1}{2}}(kr)}{i\sqrt{kr}} P_\ell^1(\cos\theta)\widehat{\boldsymbol{\theta}}\right\}\exp(-i\omega t). \tag{4}$$

$m \neq 0$ TE mode ($E_r = 0$):
$$\boldsymbol{E}(\boldsymbol{r},t) = E_0\left[\frac{J_{\ell+\frac{1}{2}}(kr)}{\sqrt{kr}}\frac{P_\ell^m(\cos\theta)}{\sin\theta}\widehat{\boldsymbol{\theta}} + \frac{J_{\ell+\frac{1}{2}}(kr)}{im\sqrt{kr}}\sin\theta\, \dot{P}_\ell^m(\cos\theta)\widehat{\boldsymbol{\varphi}}\right]\exp[i(m\varphi - \omega t)]. \tag{5}$$

$$\boldsymbol{H}(\boldsymbol{r},t) = -\frac{E_0}{\mu_0\mu(\omega)r\omega}\left\{\frac{\ell(\ell+1)J_{\ell+\frac{1}{2}}(kr)}{m\sqrt{kr}} P_\ell^m(\cos\theta)\widehat{\boldsymbol{r}} - \frac{kr\dot{J}_{\ell+\frac{1}{2}}(kr) + \tfrac{1}{2}J_{\ell+\frac{1}{2}}(kr)}{m\sqrt{kr}}\sin\theta\, \dot{P}_\ell^m(\cos\theta)\widehat{\boldsymbol{\theta}}\right.$$
$$\left. - \frac{kr\dot{J}_{\ell+\frac{1}{2}}(kr) + \tfrac{1}{2}J_{\ell+\frac{1}{2}}(kr)}{i\sqrt{kr}}\frac{P_\ell^m(\cos\theta)}{\sin\theta}\widehat{\boldsymbol{\varphi}}\right\}\exp[i(m\varphi - \omega t)]. \tag{6}$$

$m = 0$ TM mode ($H_r = 0$):

$$\boldsymbol{E}(\boldsymbol{r},t) = -\frac{H_0}{\varepsilon_0\varepsilon(\omega)r\omega}\left\{\frac{J_{\ell+\frac{1}{2}}(kr)}{i\sqrt{kr}}\left[\cot\theta\,P_\ell^1(\cos\theta) - \sin\theta\,\dot{P}_\ell^1(\cos\theta)\right]\hat{\boldsymbol{r}}\right.$$

$$\left. - \frac{kr\dot{J}_{\ell+\frac{1}{2}}(kr) + \frac{1}{2}J_{\ell+\frac{1}{2}}(kr)}{i\sqrt{kr}}P_\ell^1(\cos\theta)\hat{\boldsymbol{\theta}}\right\}\exp(-i\omega t). \quad (7)$$

$$\boldsymbol{H}(\boldsymbol{r},t) = \frac{H_0 J_{\ell+\frac{1}{2}}(kr)}{\sqrt{kr}}P_\ell^1(\cos\theta)\exp(-i\omega t)\hat{\boldsymbol{\varphi}}. \quad (8)$$

$m \neq 0$ TM mode ($H_r = 0$):

$$\boldsymbol{E}(\boldsymbol{r},t) = \frac{H_0}{\varepsilon_0\varepsilon(\omega)r\omega}\left\{\frac{\ell(\ell+1)J_{\ell+\frac{1}{2}}(kr)}{m\sqrt{kr}}P_\ell^m(\cos\theta)\hat{\boldsymbol{r}} - \frac{kr\dot{J}_{\ell+\frac{1}{2}}(kr) + \frac{1}{2}J_{\ell+\frac{1}{2}}(kr)}{m\sqrt{kr}}\sin\theta\,\dot{P}_\ell^m(\cos\theta)\hat{\boldsymbol{\theta}}\right.$$

$$\left. - \frac{kr\dot{J}_{\ell+\frac{1}{2}}(kr) + \frac{1}{2}J_{\ell+\frac{1}{2}}(kr)}{i\sqrt{kr}}\frac{P_\ell^m(\cos\theta)}{\sin\theta}\hat{\boldsymbol{\varphi}}\right\}\exp[i(m\varphi - \omega t)]. \quad (9)$$

$$\boldsymbol{H}(\boldsymbol{r},t) = H_0\left[\frac{J_{\ell+\frac{1}{2}}(kr)}{\sqrt{kr}}\frac{P_\ell^m(\cos\theta)}{\sin\theta}\hat{\boldsymbol{\theta}} + \frac{J_{\ell+\frac{1}{2}}(kr)}{im\sqrt{kr}}\sin\theta\,\dot{P}_\ell^m(\cos\theta)\hat{\boldsymbol{\varphi}}\right]\exp[i(m\varphi - \omega t)]. \quad (10)$$

In the above equations, the Bessel function $J_\nu(z)$ and its derivative with respect to $z$, $\dot{J}_\nu(z)$, could be replaced by a Bessel function of the second kind, $Y_\nu(z)$, and its derivative, $\dot{Y}_\nu(z)$, or by Hankel functions of type 1 or type 2, namely, $\mathcal{H}_\nu^{(1,2)}(z)$, and corresponding derivatives $\dot{\mathcal{H}}_\nu^{(1,2)}(z)$.

The (complex) field amplitudes are denoted by $E_0$ and $H_0$. In our spherical coordinate system, the point $\boldsymbol{r}$ is at a distance $r$ from the origin, its polar and azimuthal angles being $\theta$ and $\varphi$. The oscillation frequency is $\omega$, and the wave-number $k$ is defined as $k(\omega) = n(\omega)k_0$, where $k_0 = \omega/c$, and $n(\omega) = \sqrt{\mu(\omega)\varepsilon(\omega)}$ is the refractive index of the host medium. The integers $\ell \geq 1$, and $m$ (ranging from $-\ell$ to $+\ell$) specify the polar and azimuthal mode numbers. $P_\ell^m(\zeta)$ is an associated Legendre function, while $\dot{P}_\ell^m(\zeta)$ is its derivative with respect to $\zeta$. Note that, for a given $m$, the TM mode may be obtained from the corresponding TE mode by substituting $\boldsymbol{E}$ for $\boldsymbol{H}$, and $-\boldsymbol{H}$ for $\boldsymbol{E}$, keeping in mind that $r\omega = kr/\sqrt{\mu_0\varepsilon_0\mu(\omega)\varepsilon(\omega)}$, and that the $E/H$ amplitude ratio for each mode is always given by $\sqrt{\mu_0\mu(\omega)/\varepsilon_0\varepsilon(\omega)}$. Finally, the various Bessel functions of half-integer order are defined by the following formulas [19]:

$$J_{\ell+\frac{1}{2}}(z) = \sqrt{\frac{2}{\pi z}}\left\{\sin(z - \tfrac{1}{2}\ell\pi)\sum_{k=0}^{\lfloor\ell/2\rfloor}\frac{(-1)^k(\ell+2k)!}{(2k)!(\ell-2k)!}\left(\frac{1}{2z}\right)^{2k}\right.$$

$$\left. + \cos(z - \tfrac{1}{2}\ell\pi)\sum_{k=0}^{\lfloor(\ell-1)/2\rfloor}\frac{(-1)^k(\ell+2k+1)!}{(2k+1)!(\ell-2k-1)!}\left(\frac{1}{2z}\right)^{2k+1}\right\}. \quad (11)$$

$$Y_{\ell+\frac{1}{2}}(z) = (-1)^{\ell-1}\sqrt{\frac{2}{\pi z}}\left\{\cos(z + \tfrac{1}{2}\ell\pi)\sum_{k=0}^{\lfloor\ell/2\rfloor}\frac{(-1)^k(\ell+2k)!}{(2k)!(\ell-2k)!}\left(\frac{1}{2z}\right)^{2k}\right.$$

$$\left. - \sin(z + \tfrac{1}{2}\ell\pi)\sum_{k=0}^{\lfloor(\ell-1)/2\rfloor}\frac{(-1)^k(\ell+2k+1)!}{(2k+1)!(\ell-2k-1)!}\left(\frac{1}{2z}\right)^{2k+1}\right\}. \quad (12)$$

$$\mathcal{H}^{(1)}_{\ell+\frac{1}{2}}(z) = \sqrt{\frac{2}{\pi z}} \exp\{i[z - \tfrac{1}{2}(\ell+1)\pi]\} \sum_{k=0}^{\ell} \frac{(\ell+k)!}{k!(\ell-k)!} \left(\frac{i}{2z}\right)^k. \tag{13}$$

Note that $\sqrt{z}J_{\ell+\frac{1}{2}}(z)$ is an even function of $z$ when $\ell = 1, 3, 5, \cdots$, and an odd function when $\ell = 2, 4, 6, \cdots$. This fact will be needed later on, when we try to argue that certain branch-cuts in the complex $\omega$-plane are inconsequential. Also, the following alternative representation of Bessel functions of the first kind, order $\nu$, will be found useful:

$$J_\nu(z) = (z/2)^\nu \sum_{k=0}^{\infty} \frac{(-1)^k (z/2)^{2k}}{k!\,\Gamma(\nu+k+1)}. \tag{14}$$

Given that $\nu = \ell + \frac{1}{2} \geq 3/2$ for spherical harmonics, Eq.(14) reveals that $J_{\ell+\frac{1}{2}}(z)/z \to 0$ when $z \to 0$.

Consider now a solid dielectric sphere of radius $R$, relative permeability $\mu(\omega)$, and relative permittivity $\varepsilon(\omega)$. Inside the particle, the radial dependence of the TE mode is governed by a Bessel function of the first kind, $E_0 J_{\ell+\frac{1}{2}}(kr)$, and its derivative. The refractive index of the spherical particle being $n(\omega) = \sqrt{\mu(\omega)\varepsilon(\omega)}$, the corresponding wave-number inside the particle is $k(\omega) = n(\omega)k_0 = n(\omega)\omega/c$. The particle is surrounded by free space, which is host to an *outgoing* spherical harmonic whose radial dependence is governed by a type 1 Hankel function, $E_1 \mathcal{H}^{(1)}_{\ell+\frac{1}{2}}(k_0 r)$, and its derivative. Invoking the Bessel function identity $zJ'_\nu(z) = \nu J_\nu(z) - zJ_{\nu+1}(z)$—which applies to $Y_\nu(z)$ and $\mathcal{H}^{(1,2)}_\nu(z)$ as well—we find, upon matching the boundary conditions at $r = R$, that the following two equations must be simultaneously satisfied:

$$\frac{E_0 J_{\ell+\frac{1}{2}}(nk_0 R)}{\sqrt{nk_0 R}} = \frac{E_1 \mathcal{H}^{(1)}_{\ell+\frac{1}{2}}(k_0 R)}{\sqrt{k_0 R}}, \tag{15a}$$

$$\frac{E_0[(\ell+1)J_{\ell+\frac{1}{2}}(nk_0 R) - nk_0 R J_{\ell+3/2}(nk_0 R)]}{\mu(\omega)\sqrt{nk_0 R}} = \frac{E_1[(\ell+1)\mathcal{H}^{(1)}_{\ell+\frac{1}{2}}(k_0 R) - k_0 R \mathcal{H}^{(1)}_{\ell+3/2}(k_0 R)]}{\sqrt{k_0 R}}. \tag{15b}$$

Streamlining the above equations, we arrive at

$$\begin{bmatrix} J_{\ell+\frac{1}{2}}(nk_0 R) & -\sqrt{n}\mathcal{H}^{(1)}_{\ell+\frac{1}{2}}(k_0 R) \\ (\ell+1)J_{\ell+\frac{1}{2}}(nk_0 R) - nk_0 R J_{\ell+3/2}(nk_0 R) & -\mu\sqrt{n}[(\ell+1)\mathcal{H}^{(1)}_{\ell+\frac{1}{2}}(k_0 R) - k_0 R \mathcal{H}^{(1)}_{\ell+3/2}(k_0 R)] \end{bmatrix} \begin{bmatrix} E_0 \\ E_1 \end{bmatrix} = 0. \tag{16}$$

A non-trivial solution for $E_0$ and $E_1$ thus exists if and only if the determinant of the coefficient matrix in Eq.(16) vanishes, that is,

$$F(\omega) = nk_0 R \mathcal{H}^{(1)}_{\ell+\frac{1}{2}}(k_0 R) J_{\ell+3/2}(nk_0 R) + \left[(\mu-1)(\ell+1)\mathcal{H}^{(1)}_{\ell+\frac{1}{2}}(k_0 R) - \mu k_0 R \mathcal{H}^{(1)}_{\ell+3/2}(k_0 R)\right] J_{\ell+\frac{1}{2}}(nk_0 R) = 0. \tag{17}$$

This is the characteristic equation for leaky TE modes, whose solutions comprise the entire set of leaky frequencies $\omega_q$. (The index $q$ is used here to enumerate the various leaky-mode frequencies.) For TM modes, $\mu(\omega)$ in Eq.(17) must be replaced by $\varepsilon(\omega)$.

Equation (17) must be solved numerically for complex frequencies $\omega_q$; these being characteristic frequencies of the spherical particle's leaky modes, one expects (on physical grounds) to find all the roots $\omega_q$ of $F(\omega)$ in the lower-half of the complex plane. Note that $\sqrt{n}F(\omega)$ is an even function of $n$ when $\ell = 1, 3, 5, \cdots$, and an odd function when $\ell = 2, 4, 6, \cdots$. This is because successive Bessel functions $J_{\ell+\frac{1}{2}}$ and $J_{\ell+3/2}$ alternate between odd and even parities. Note also that $F(\omega)$ vanishes at the zeros of $n(\omega)$, that is, $F(\Omega_1) = F(\Omega_2) = 0$; see Eq.(2b). Nevertheless, $\Omega_1$ and $\Omega_2$ do *not* represent leaky-mode frequencies, because setting $n(\Omega_{1,2}) = 0$ in Eqs.(3)-(10) extinguishes the EM field throughout the dielectric sphere. At the poles of $n(\omega)$, namely, $\omega = \Omega_3$ and $\omega = \Omega_4$ given by Eq.(2c), the function $F(\omega)$ is undefined, but an arbitrarily small circle centered at $\Omega_3$ (or $\Omega_4$) can be shown to contain an infinite number of the zeros of $F(\omega)$. One could argue that, throughout the dielectric sphere, the EM fields associated with the $\Omega_{3,4}$ frequencies should be negligible, although the mathematical reasoning behind this argument is not straightforward. Finally, when $\omega \to 0$, $F(\omega)$ approaches a constant (see the Appendix), and when $|\omega| \to \infty$, $\mu(\omega) \to 1 - (\omega_{pm}/\omega)^2$ and $\varepsilon(\omega) \to 1 - (\omega_{pe}/\omega)^2$, thus allowing the asymptotic behavior of $F(\omega)$ to be determined from Eqs.(11) and (13).

Our goal is to express an initial field distribution inside the spherical particle (e.g., one of the spherical harmonic waveforms given by Eqs.(3)-(10), which oscillate at a real-valued frequency $\omega_0$) as a superposition of leaky modes, each having its own complex frequency $\omega_q$. To this end, we must form a meromorphic function $G(\omega)$ incorporating the following features:

i) The function $F(\omega)$ of Eq.(17) appears in the denominator of $G(\omega)$, thus causing the zeros of $F(\omega)$ to act as poles for $G(\omega)$.

ii) A desired initial waveform, say, $J_{\ell+½}[\omega n(\omega) r/c]$, appearing in the numerator of $G(\omega)$.

iii) The real-valued frequency $\omega_0$ associated with the initial waveform acting as a pole for $G(\omega)$.

iv) In the limit when $|\omega| \to \infty$, $G(\omega) \to 0$ exponentially, so that $\oint G(\omega) d\omega$ over a circle of large radius $R_c$ would vanish.

A simple (although by no means the only) such function is

$$G(\omega) = \frac{\sqrt{\omega} \exp(iR\omega/c) J_{\ell+½}(kr)}{(\omega - \omega_0) F(\omega)}. \tag{18}$$

With reference to Eq.(11), note that the pre-factor $1/\sqrt{n}$ of the Bessel function in the numerator of $G(\omega)$ cancels the corresponding pre-factor that accompanies the denominator. The remaining part of the Bessel function in the numerator will then have the same parity with respect to $n(\omega)$ as the function that appears in the denominator. Consequently, switching the sign of $n(\omega)$ does *not* alter $G(\omega)$, indicating that the branch-cuts associated with $n(\omega)$ in the complex $\omega$-plane do *not* introduce discontinuities into $G(\omega)$. The presence of $\sqrt{\omega} \exp(iR\omega/c)$ in the numerator of $G(\omega)$ is intended to eliminate certain undesirable features of the Hankel functions appearing in the denominator. The function $G(\omega)$ is thus analytic everywhere except at its poles, where its denominator vanishes. The poles, of course, consist of $\omega = \omega_0$, which is the frequency of the initial EM field residing inside the spherical particle at $t = 0$, and $\omega = \omega_q$, which are the leaky-mode frequencies found by solving Eq.(17)—or its TM mode counterpart. The zeros of the refractive index $n(\omega)$, namely, $\omega = \Omega_{1,2}$, do not become poles of $G(\omega)$, because the numerator of $G(\Omega_{1,2})$ also equals zero. At the poles $\omega = \Omega_{3,4}$ of the refractive index, $G(\omega)$ is undefined, but it is well-behaved in the sense that the integral of $G(\omega)$ around a small circle centered at $\omega = \Omega_{3,4}$, whose radius passes between consecutive poles, approaches zero as the radius of the circle goes to zero. For this reason, one can invoke Cauchy's theorem in order to construct a leaky-mode expansion for the dielectric sphere, even though the integrand, $G(\omega)$, has non-isolated singularities at $\Omega_{3,4}$ (and also when $\omega \to \infty$). These mathematical details will be addressed in a forthcoming paper.

In the limit $|\omega| \to \infty$, where $\mu(\omega) \to 1 - (\omega_{pm}/\omega)^2$ and $\varepsilon(\omega) \to 1 - (\omega_{pe}/\omega)^2$, we find that $G(\omega)$ approaches zero exponentially. Thus, the vanishing of $\oint G(\omega)d\omega$ around a circle of large radius $R_c$ ensures that all the residues of $G(\omega)$ add up to zero, that is,

$$\frac{\sqrt{\omega_0}\exp(iR\omega_0/c)J_{\ell+½}[\omega_0 n(\omega_0)r/c]}{F(\omega_0)} + \sum_q \frac{\sqrt{\omega_q}\exp(iR\omega_q/c)J_{\ell+½}[\omega_q n(\omega_q)r/c]}{(\omega_q - \omega_0)F'(\omega_q)} = 0. \quad (19)$$

The initial field distribution $J_{\ell+½}[\omega_0 n(\omega_0)r/c]$ may thus be expanded as the following superposition of all the leaky modes:

$$J_{\ell+½}[\omega_0 n(\omega_0)r/c] = \sum_q \frac{(\omega_q/\omega_0)^{½}\exp[iR(\omega_q - \omega_0)/c]F(\omega_0)}{(\omega_0 - \omega_q)F'(\omega_q)} \times J_{\ell+½}[\omega_q n(\omega_q)r/c]. \quad (20)$$

To incorporate into the initial distribution the denominator $\sqrt{kr}$, which accompanies all the field components in Eqs.(3)-(10), we modify Eq.(20) — albeit trivially — as follows:

$$\frac{J_{\ell+½}[\omega_0 n(\omega_0)r/c]}{\sqrt{\omega_0 n(\omega_0)r/c}} = \sum_q \frac{(\omega_q/\omega_0)[n(\omega_q)/n(\omega_0)]^{½}\exp[iR(\omega_q - \omega_0)/c]F(\omega_0)}{(\omega_0 - \omega_q)F'(\omega_q)} \times \frac{J_{\ell+½}[\omega_q n(\omega_q)r/c]}{\sqrt{\omega_q n(\omega_q)r/c}}. \quad (21)$$

The above formula is a central result of the present paper, indicating that a general EM field distribution excited from outside the cavity can be represented by a superposition of leaky modes. Indeed, upon termination of the external excitation, the field that remains within the cavity is, in general, a superposition of functions similar to that appearing on the left-hand-side of Eq.(21), with the spectral weight associated with each such function depending on its oscillation frequency $\omega_0$. Thus, with the important caveat discussed in the following paragraph, Eq.(21) provides an explicit formula for computing the leaky-mode expansion coefficients corresponding to the post-excitation evolution of the intra-cavity field.

Without going into details, it must be pointed out that the argument for the vanishing of the contour integral around a large circle in the $\omega$-plane contains a couple of subtleties. One is that the integration contour must pass between the poles that represent the very resonances used for the expansion. While the straightforward reasoning about the exponential decay of the integrand cannot be applied to such a portion of the integral, it can be shown that its contribution does indeed vanish in the limit $R_c \to \infty$ if our choice for $G(\omega)$ as given by Eq.(18) is somewhat modified in such a way as to accelerate its approach to zero when $|\omega| \to \infty$. The second issue is that, besides $\omega \to \infty$ being an accumulation point for the singularities of $G(\omega)$, there exist other such points, namely, the poles $\Omega_{3,4}$ of the Lorentzian refractive index; see Fig.1. In this case, it can be shown that the requirements for the series convergence are less restrictive than those pertaining to $\omega \to \infty$. In fact, one can introduce additional poles into $G(\omega)$ by multiplying its denominator with $(\omega - \Omega_3)(\omega - \Omega_4)$ and still obtain a convergent series. These convergence issues are brought about by the dispersion properties of the refractive index together with the fact that $n(\omega) \to 1$ when $|\omega| \to \infty$, issues that, to the best of our knowledge, have not been discussed in the existing literature concerning leaky-mode expansion of dispersive optical cavities. Unfortunately, a detailed exposition of the convergence proof is beyond the scope of the present paper and must be presented elsewhere. The bottom line is that the convergence of the series can be guaranteed if the leaky-mode expansion coefficients in Eq.(21) are multiplied by the additional factor $(\omega_0 - \Omega_3)(\omega_0 - \Omega_4)/[(\omega_q - \Omega_3)(\omega_q - \Omega_4)]$.

Taking advantage of the flexibility of $G(\omega)$, we now extend the same treatment to the remaining components of the EM field. For instance, if we choose

$$G(\omega) = \frac{\sqrt{\omega} \exp(iR\omega/c) J_{\ell+\frac{1}{2}}(kr)}{\omega(\omega - \omega_0)(\omega - \Omega_3)(\omega - \Omega_4)\mu(\omega)F(\omega)}, \qquad (22)$$

then $G(\omega) \to 0$ exponentially in the limit when $|\omega| \to \infty$, resulting in a vanishing integral around the circle of large radius $R_c$ in the $\omega$-plane. We thus arrive at an alternative form of Eq.(21), which is useful for expanding the field component $H_r$ appearing in Eqs.(4) and (6), that is,

$$\frac{J_{\ell+\frac{1}{2}}[\omega_0 n(\omega_0)r/c]}{\mu(\omega_0)r\omega_0\sqrt{\omega_0 n(\omega_0)r/c}} = \sum_q \frac{(\omega_q/\omega_0)[n(\omega_q)/n(\omega_0)]^{\frac{1}{2}} \exp[iR(\omega_q - \omega_0)/c]F(\omega_0)}{(\omega_0 - \omega_q)F'(\omega_q)}$$

$$\times \frac{(\omega_0 - \Omega_3)(\omega_0 - \Omega_4)}{(\omega_q - \Omega_3)(\omega_q - \Omega_4)} \times \frac{J_{\ell+\frac{1}{2}}[\omega_q n(\omega_q)r/c]}{\mu(\omega_q)r\omega_q[\omega_q n(\omega_q)r/c]^{\frac{1}{2}}}. \qquad (23)$$

Similarly, if we choose

$$G(\omega) = \frac{\sqrt{\omega} \exp(iR\omega/c)[krj_{\ell+\frac{1}{2}}(kr) + \frac{1}{2}J_{\ell+\frac{1}{2}}(kr)]}{\omega(\omega - \omega_0)(\omega - \Omega_3)(\omega - \Omega_4)\mu(\omega)F(\omega)}, \qquad (24)$$

it continues to be meromorphic (i.e., free of branch-cuts), and will have a vanishing integral over a large circle of radius $R_c$ in the limit when $R_c \to \infty$. The relevant expansion of the field components $H_\theta$ and $H_\varphi$ appearing in Eqs.(4) and (6) will then be obtained from $G(\omega)$ of Eq.(24).

In this way, one can expand into a superposition of leaky modes the various $E$- and $H$-field components that comprise an initial distribution. It will then be possible to follow each leaky mode as its phase evolves while its amplitude decays with the passage of time.

As for the beam that leaks out of the cavity and into the free-space region $r > R$, it can be shown that the fields *grow* exponentially along the radial direction, but of course this exponential growth terminates at $r = ct$, where the leaked beam meets up with the tail end of the beam that was originally reflected from the surface of the sphere (i.e., prior to the abrupt termination of the incident beam at $t = 0$). The EM energy in the region $R < r < ct$ is just the energy that has leaked out of the spherical cavity, with the exponential decline of the field amplitude in time compensating for the expansion of the region "illuminated" by the leaked beam.

Before concluding this section, a note concerning over-completeness might be in order. It is known that resonant modes are subject to sum rules which make it possible to create nontrivial linear combinations that sum-up to zero [20,21]. Our method also allows derivation of such sum rules. To this end it is sufficient to remove the factor $\omega/(\omega - \omega_0)$ from the function $G(\omega)$. This will not modify the asymptotic behavior at infinity, but it eliminates the contribution of the pole at $\omega_0$, thus giving rise to an over-completeness relation.

**4. Numerical results**. As pointed out earlier, the zeros $\omega_q$ of the characteristic function $F(\omega)$ appearing in Eq.(17) must be confined to the lower-half of the complex $\omega$-plane. This is because, when the incident beam is removed, the time-dependence factor $\exp(-i\omega_q t)$ of the corresponding leaky modes inside and outside the cavity can only decrease with time. Also, considering that $\varepsilon(-\omega_q^*) = \varepsilon^*(\omega_q)$, and $\mu(-\omega_q^*) = \mu^*(\omega_q)$, and $n(-\omega_q^*) = n^*(\omega_q)$, the zeros of $F(\omega)$ always appear in pairs such as $\omega_q$ and $-\omega_q^*$. Consequently, leaky frequencies appear in the third and fourth quadrants of the $\omega$-plane as mirror images of each other.

Trivial leaky modes occur at $\omega_q = \Omega_{1m}$ and $\Omega_{1e}$ (with their twins occurring at $-\omega_q^* = \Omega_{2m}$ and $\Omega_{2e}$), where $n(\Omega_{1,2}) = 0$. Substitution into Eqs.(3)-(10) reveals that, for these trivial leaky modes, which are associated with the zeros of the refractive index $n(\omega)$, both $E$ and $H$ fields inside and outside the cavity vanish. Finally, referring to the complex $\omega$-plane of Fig.1, note that when $\omega$ crosses (i.e., moves from immediately above to immediately below) one of the branch-cuts, $n(\omega)$ gets multiplied by $-1$, which causes $F(\omega)$ of Eq.(17) to be multiplied by $\pm i$ (depending on the value of $\ell$ being even or odd).

The contour plots in Fig.2 show, within two segments of the fourth quadrant of the $\omega$-plane, the zeros of $\text{Re}[F(\omega)]$ in red (solid) lines and the zeros of $\text{Im}[F(\omega)]$ in blue (dashed) lines. Both $\text{Re}(\omega)$ and $\text{Im}(\omega)$ are normalized by the (arbitrarily chosen) reference frequency $\omega_{\text{ref}} = 1.216 \times 10^{15}$ rad/sec, which corresponds to the free-space wavelength $\lambda_{\text{ref}} = 1.55\,\mu m$. The chosen value of $\ell$ for the plots of Fig.2 is 10, the dielectric sphere has radius $R = 1.55\,\mu m$, permeability $\mu(\omega) = 1.0$, and the refractive index, $n(\omega) = \sqrt{\varepsilon(\omega)}$, is governed by a single Lorentz oscillator having $\omega_r = 2\omega_{\text{ref}}$, $\omega_p = 5\omega_{\text{ref}}$, and $\gamma = 0.02\omega_{\text{ref}}$. The 4$^{\text{th}}$ quadrant pole and zero of $n(\omega)$ are thus located at $\Omega_{3e} \cong (2.0 - 0.01i)\omega_{\text{ref}}$ and $\Omega_{1e} \cong (5.385 - 0.01i)\omega_{\text{ref}}$, respectively. The parameter values chosen here do not necessarily represent a realistic cavity such as a fused silica micro-sphere. Nevertheless, we have chosen these values with the following illustration in mind. Despite being artificial, they preserve the "topology" of the resonant pole distribution in the $\omega$-plane, while allowing a reasonable visualization. The small size of the cavity, together with a strongly lossy and dispersive medium, effectively isolates the important features that we would like to show.

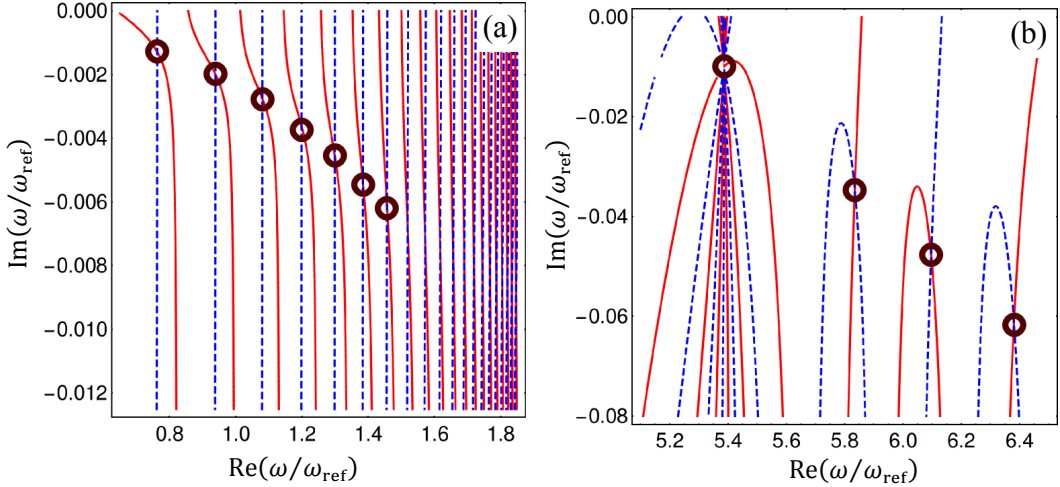

**Fig.2**. Contours in the complex $\omega$-plane representing regions where $\text{Re}[F(\omega)] = 0$ (solid red lines) and $\text{Im}[F(\omega)] = 0$ (dashed blue lines). The real and imaginary axes are normalized by the reference frequency $\omega_{\text{ref}} = 1.216 \times 10^{15}$ rad/sec. Where a solid red and a dashed blue curve cross, $F(\omega)$ vanishes; these crossing points (some of them marked with small circles) correspond to TE leaky-mode frequencies $\omega_q = \omega'_q + i\omega''_q$ of the spherical cavity at the chosen value of $\ell = 10$. The spherical particle has radius $R = 1.55\,\mu m$, permeability $\mu(\omega) = 1.0$, and refractive index $n(\omega) = \sqrt{\varepsilon(\omega)}$ governed by a single Lorentz oscillator. The 4$^{\text{th}}$ quadrant pole and zero of $n(\omega)$ are, respectively, at $\Omega_{3e} \cong (2.0 - 0.01i)\omega_{\text{ref}}$ and $\Omega_{1e} \cong (5.385 - 0.01i)\omega_{\text{ref}}$.

The points where the contours depicted in Fig.2 cross each other—several crossing points are circled in the plot—represent the zeros of $F(\omega)$, which we have denoted by $\omega_q = \omega'_q + i\omega''_q$ and referred to as leaky-mode frequencies. The region of the $\omega$-plane depicted in Fig.2(a) contains the 4$^{\text{th}}$ quadrant leaky-mode frequencies to the left of $\Omega_{3e}$; a large number of such frequencies are seen to accumulate in the vicinity of $\omega = \Omega_{3e}$, where the coupling of the incident light to the cavity is weak, and the damping within the sphere is dominated by absorption losses.

The region of the $\omega$-plane depicted in Fig.2(b) contains the 4th quadrant leaky-mode frequencies to the right of $\Omega_{1e}$. The imaginary part $\omega_q''$ of these leaky frequencies is seen to acquire large negative values as the corresponding real part $\omega_q'$ increases. No leaky frequencies were found in the upper-half of the $\omega$-plane, nor were there any in the strip between $\Omega_{1e}$ and $\Omega_{3e}$. As mentioned earlier, symmetry considerations ensure that the poles in the third and fourth quadrants are mirror-images of each other. As will be seen shortly, when the dielectric sphere is illuminated with a real-valued excitation frequency $\omega$, resonances occur in the vicinity of $\omega = \omega_q'$, i.e., at and around the real parts of the various leaky mode frequencies.

Note that the leftmost zero-crossing shown in Fig.2(b) represents a zero of the refractive index $n(\omega)$, which has multiplicity equal to the order of the Bessel function associated with the modal field. However, this zero of the function $F(\omega)$ is cancelled out by the numerator of $G(\omega)$, as can be readily seen by expanding in the vicinity of the complex zero of $n(\omega)$. As such, the leftmost zero-crossing in Fig.2(b) does not contribute to the leaky-mode expansion.

To investigate the resonant behavior of the dielectric sphere described in conjunction with Fig.2, we pick a real-valued frequency $\omega$, then select a mode consisting of incoming and outgoing Hankel functions outside the sphere, matched to a Bessel function of the first kind residing inside. The resulting equations do not depend on the azimuthal mode number $m$, which indicates that, for a given integer $\ell$, the modes associated with all values of $m$ between $-\ell$ and $\ell$ are degenerate. Figure 3 shows the computed amplitude ratio of the $E$-field inside the sphere to the incident $E$-field, plotted versus the normalized excitation frequency $\omega/\omega_{\text{ref}}$. Here the $E$-field amplitude is defined as the magnitude of $E_0$ in Eq.(5). As before, $R = 1.55\ \mu m$, $\mu(\omega) = 1.0$, $\varepsilon(\omega)$ follows a single Lorentz oscillator model ($\omega_r = 2\omega_{\text{ref}}, \omega_p = 5\omega_{\text{ref}}, \gamma = 0.02\omega_{\text{ref}}$), and the selected TE mode has $\ell = 10$. In the interval $[\Omega_{3e}, \Omega_{1e}]$ between the pole and zero of the refractive index (see Fig.1), the field amplitude inside the cavity is seen to be vanishingly small. Outside this "forbidden" zone, the field has resonance peaks at specific frequencies, and the ratio $E_{\text{inside}}/E_{\text{incident}}$ can vary significantly between adjacent peaks and valleys. For the chosen set of parameters in Fig.3, the minimum resonance frequency occurs at $\omega \cong 0.76253\omega_{\text{ref}}$.

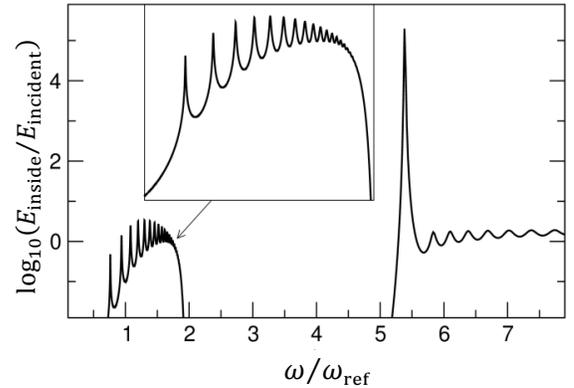

**Fig.3**. Logarithmic plot of the ratio of the $E$-field inside the sphere to the incident $E$-field, for a dielectric sphere of radius $R = 1.55\ \mu m$ at $\ell = 10$. The horizontal axis represents the normalized excitation frequency $\omega/\omega_{\text{ref}}$. The refractive index $n(\omega) = \sqrt{\varepsilon(\omega)}$ of the spherical particle is governed by a single Lorentz oscillator. The 4th quadrant pole and zero of $n(\omega)$ are at $\Omega_{3e} \cong (2.0 - 0.01i)\omega_{\text{ref}}$ and $\Omega_{1e} \cong (5.385 - 0.01i)\omega_{\text{ref}}$, respectively. The EM field hardly penetrates the dielectric sphere in the frequency interval between the pole and zero of $n(\omega)$. Outside this "forbidden" interval, the $E$-field amplitude ratio exhibits sharp peaks at certain frequencies, which is indicative of resonant behavior.

A comparison of Fig.2 with Fig.3 reveals a close relationship between the leaky mode frequencies and the resonances of the dielectric sphere. Resonances occur at or near the (real-valued) frequencies $\omega = \omega_q'$, and the height and width of a resonance line are, by and large, determined by the decay rate $\omega_q''$ of the corresponding leaky mode—unless the leaky mode frequency happens to be so close to the pole(s) of the refractive index $n(\omega)$ that the strong absorption within the medium would suppress the resonance. It must be emphasized that the presence of a gap in the frequency domain (such as that between $\omega = \text{Re}(\Omega_{3e})$ and $\omega =$

Re($\Omega_{1e}$) in the present example) should not prevent the leaky modes from forming a complete basis. This is because one expects, on physical grounds, that the ensemble of leaky modes would carry all the spatial frequencies needed to capture the various features of an arbitrary initial EM field distribution.

For leaky TE modes, the radial dependence of the $E$-field inside and outside the dielectric sphere are given by $E_0 J_{\ell+\frac{1}{2}}(kr)/\sqrt{kr}$ and $E_1 \mathcal{H}^{(1)}_{\ell+\frac{1}{2}}(k_o r)/\sqrt{k_o r}$, respectively. Here $k(\omega) = n(\omega)\omega/c$ and $k_o(\omega) = \omega/c$. Plots of the $E$-field amplitude for several leaky $\ell = 10$ TE modes of a sphere of radius $R = 1.55\,\mu m$ are shown in Fig.4. The refractive index of the dielectric material at the reference frequency $\omega_{\text{ref}} = 1.216 \times 10^{15}$ rad/sec is $n(\omega_{\text{ref}}) = 3.055 + 0.0091i$. The fields are plotted as functions of the normalized radial coordinate $r/R$, with frames (a) and (b) depicting the real and imaginary components of the $E$-field. The solid (black) curve, the dashed (red) curve, and the dash-dotted (blue) curve correspond to $\omega/\omega_{\text{ref}} = 0.76253 + 0.00128i$, $0.938779 + 0.00199i$, and $1.08039 + 0.00275i$, respectively.

Figure 5 provides a comparison between the $E$-field inside a spherical cavity and its expansion in terms of the leaky-modes ($R = 1.55\,\mu m$, $\ell = 10$ TE mode, $\omega = 1.8\omega_{\text{ref}}$). The solid black and solid red lines show, respectively, the real and imaginary parts of the target solution, $E_{\text{inside}}(r)$, whereas the symbols superposed on these solid lines represent the leaky-mode expansion, $\sum_q E_{\text{leaky}}(r)$, of the target function composed of 100 terms. The convergence is seen to be rather poor near the surface of the sphere ($0.9 < r/R < 1.0$).

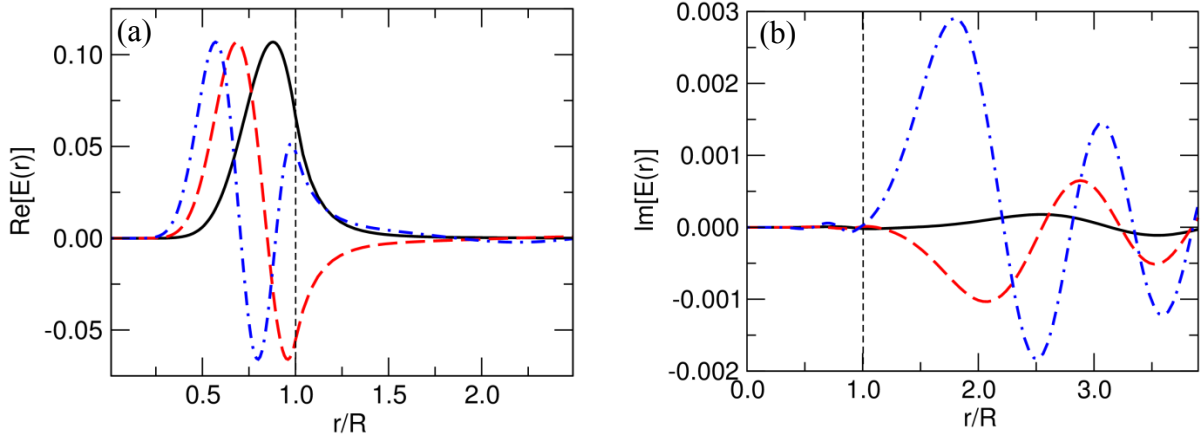

**Fig.4**. The $E$-field amplitude inside and outside a solid dielectric sphere of radius $R = 1.55\,\mu m$, plotted versus the normalized radial coordinate $r/R$ for three $\ell = 10$ TE modes. The real part of the field is shown in (a), while its imaginary part appears in (b). The solid black, dashed red, and dash-dotted blue curves correspond, respectively, to $\omega/\omega_{\text{ref}} = 0.76253 + 0.00128i$, $0.938779 + 0.00199i$, and $1.08039 + 0.00275i$ ($\omega_{\text{ref}} = 1.216 \times 10^{15}$ rad/sec). The particle, whose permeability is $\mu(\omega) = 1.0$, has a refractive index $n(\omega) = \sqrt{\varepsilon(\omega)}$ governed by a single Lorentz oscillator ($\omega_r = 2\omega_{\text{ref}}$, $\omega_p = 5\omega_{\text{ref}}$, and $\gamma = 0.02\omega_{\text{ref}}$).

The error of the leaky-mode expansion depicted in Fig.5 (superposed on the function being expanded) illustrates that the gap between the expansion and its target function, while indiscernible at small radii, grows in the vicinity of the boundary of the cavity. This behavior is generic, and a manifestation of the fact that our leaky-mode expansion converges rather slowly. In fact, adding hundreds or even thousands of terms to the expansion only results in a minuscule reduction in the residual error. We have traced this behavior to the fact that the terms in the expansion do not enter their asymptotic regime until their count is on the order of $10^{10}$. The practical consequence here is that, while a good approximation can be achieved with a fairly small number of terms, suppressing the error below a few parts in a thousand becomes utterly impractical. That being said, one should keep in mind that the expansion error is due primarily to those basis functions that decay rapidly upon termination of the excitation. In other words, due to the large imaginary parts $\omega_q''$ of their eigen-frequencies, the contribution of high-order leaky modes will disappear almost instantly once the excitation is terminated. If, for some applications, accuracy beyond a few parts in a thousand turns out to be necessary, it is worth noting that, with the asymptotic information about convergence rates that can easily be determined for these series, it is highly likely that convergence accelerating re-summation methods can be deployed.

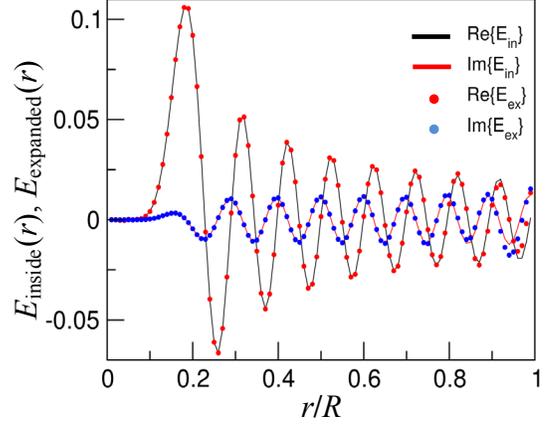

**Fig.5**. Comparison of the $E$-field inside a spherical cavity with its expansion as a superposition of leaky modes ($R = 1.55\ \mu m$, $\ell = 10$ TE mode, $\omega = 1.8\omega_{\text{ref}}$). The real and imaginary parts of $E_{\text{inside}}(r)$ are shown as solid lines—black and red (grey), respectively. The superposed symbols (i.e., small solid circles) represent the result of leaky-mode expansion, $\sum_q E_{\text{leaky}}(r)$, composed of 100 terms.

For the remaining set of figures, we shall ignore the dispersive nature of the dielectric host and simply assume that $\mu(\omega) = 1.0$ and $\varepsilon(\omega) = 2.25$ at and around the reference frequency $\omega_{\text{ref}} = 1.216 \times 10^{15}$ rad/sec (corresponding to the vacuum wavelength $\lambda_{\text{ref}} = 1.55\ \mu m$). This is tantamount to confining the frequency range of interest to $\Omega_{1m} \ll \omega \ll \Omega_{3e}$. Unlike the previous example in which the parameter selection was driven by the visualization needs, the parameter values in the following examples are comparable to those found in actual experiments [1,2].

Figure 6 shows the resonances of a dielectric sphere of radius $R = 50\lambda_{\text{ref}}$ and refractive index $n = 1.5$ for the $\ell = 340$ TE and TM modes. The contours of real and imaginary parts of the characteristic equation $F(\omega) = 0$ have been plotted in the $\omega$-plane, as was done for a different set of parameters in Fig.2. Where the contours cross each other, the function $F(\omega)$ vanishes, indicating the existence of a leaky mode at the crossing frequency $\omega_q = \omega_q' + i\omega_q''$. The ratio $|\omega_q'/\omega_q''|$ is a measure of the $Q$-factor of the spherical cavity at (or near) the excitation frequency $\omega = \omega_q'$.

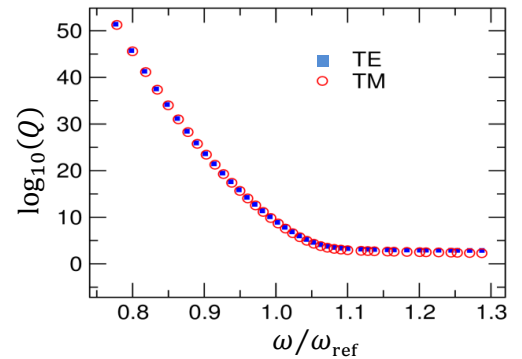

**Fig. 6**. Computed $Q$-factor versus the resonance frequency for a solid dielectric sphere ($R = 77.5\ \mu m$, $\mu = 1$, $n = 1.5$) in the vicinity of $\omega_{\text{ref}} = 1.216 \times 10^{15}$ rad/sec. The leaky frequencies $\omega_q = \omega_q' + i\omega_q''$ are solutions of $F(\omega) = 0$, which have been found numerically. The ratio $|\omega_q'/\omega_q''|$ is used as a measure of the cavity $Q$-factor at the excitation frequency $\omega = \omega_q'$. Shown are computed $Q$-factors for both TE and TM modes (solid blue squares for TE, open red circles for TM) at several resonance frequencies of the dielectric sphere corresponding to $\ell = 340$.

Shown in Fig.6 are the computed $Q$-factors of the spherical cavity for both TE and TM modes at the various resonance frequencies corresponding to $\ell = 340$. (Note that the characteristic equation does not depend on $m$, which indicates that, for a given integer $\ell$, the modes associated with all $m$ between $-\ell$ and $\ell$ are degenerate.) The lowest resonance frequency occurs at $\omega \cong 0.78\omega_{\text{ref}}$. The large values of $Q$ seen in Fig.6 are a consequence of the fact that the refractive index $n$ is assumed to be purely real; later, when absorption is incorporated into the model via the imaginary part of $n$, the $Q$-factors will drop to more reasonable values.

The direct method of determining the resonances of the spherical cavity involves the computation of the amplitude ratio $E_{\text{inside}}/E_{\text{incident}}$ for an incident Hankel function of type 2 (incoming wave) and a fixed mode number $\ell$. (As pointed out earlier, the amplitude of each $E$-field is defined as the magnitude of the corresponding $E_0$ in Eq.(5), with the radial dependence of the inside field being given in terms of the Bessel function $J_{\ell+\frac{1}{2}}(kr)$, while that of the incident field outside the sphere involves the Hankel function $\mathcal{H}^{(2)}_\nu(k_0 r)$.) Once again, the results are independent of the azimuthal mode number $m$, as the modes associated with $m = -\ell$ to $\ell$ are all degenerate. Figure 7 shows plots of $E_{\text{inside}}/E_{\text{incident}}$ for the spherical cavity of radius $R = 50\lambda_{\text{ref}}$, refractive index $n = 1.5$, and mode number $\ell = 340$, at and around $\omega_{\text{ref}} = 1.216 \times 10^{15}$ rad/sec; the results for both TE and TM modes are presented in the figure. The resonances are seen to be strong, with narrow linewidths.

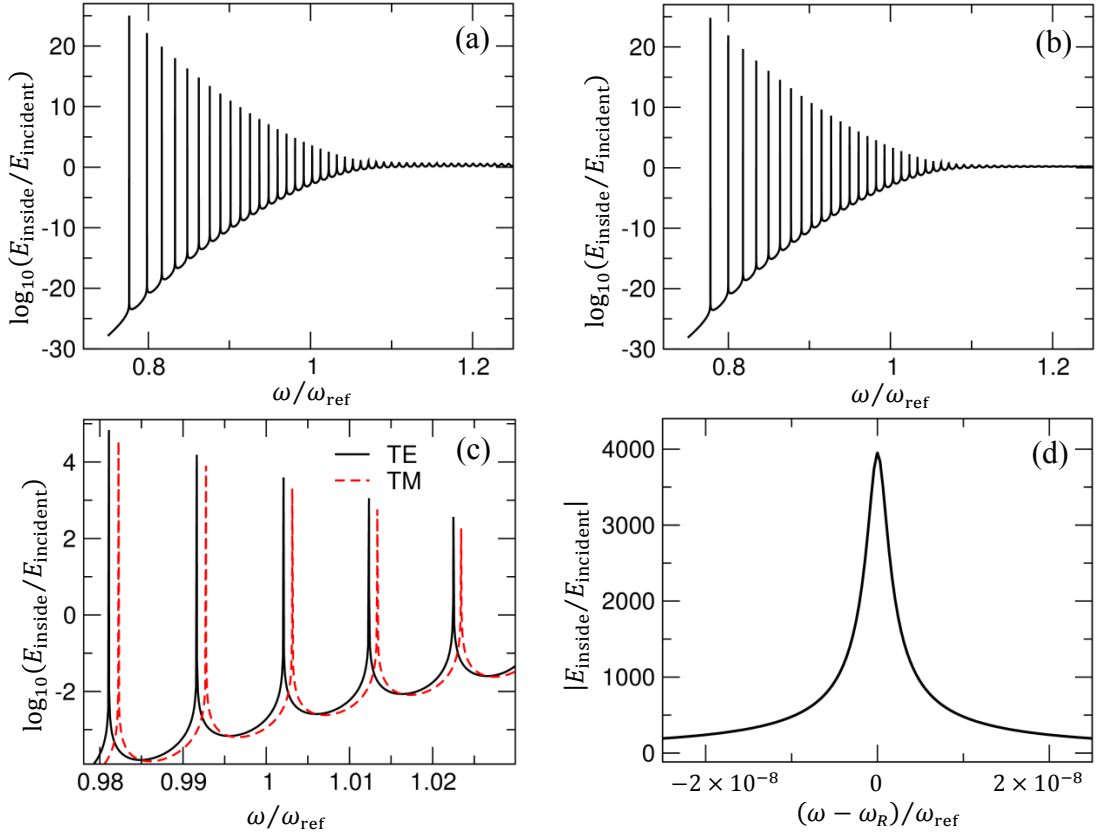

**Fig. 7**. Plots of the amplitude ratio of the $E$-field inside the dielectric sphere ($R = 77.5 \, \mu m, \mu = 1, n = 1.5$) to the incident $E$-field for the $\ell = 340$ spherical harmonic. The horizontal axis represents the excitation frequency $\omega$ normalized by $\omega_{\text{ref}} = 1.216 \times 10^{15}$ rad/sec. (a) TE mode. (b) TM mode. Note that the cutoff frequency for both modes is $\omega \cong 0.78\omega_{\text{ref}}$, below which no resonances are excited. Above the cutoff, in between adjacent resonances, the field amplitude inside the cavity drops to exceedingly small values. The occurrence of extremely large resonance peaks in these plots is due to the assumed value of the refractive index $n$ being purely real. (c) Close-up view of the resonance lines of the glass ball for the $\ell = 340$ spherical harmonic, showing the TM resonances (dashed red lines) being slightly shifted away from the TE resonances (solid black lines). (d) Magnified view of an individual TE resonance line centered at $\omega_R = 1.00207\omega_{\text{ref}}$.

Outside the resonance peaks and especially at lower frequencies, it is seen that the coupling of the incident beam to the cavity is extremely weak. The TE and TM modes are quite similar in their coupling efficiencies and resonant line-shapes, their major difference being the slight shift of TM resonances toward higher frequencies, as can be seen in Fig.7(c). Figure 7(d) is a magnified view of the line-shape for a single TE resonant line centered at $\omega = 1.00207\omega_{\text{ref}}$.

To gain an appreciation for the effect of the mode number $\ell$ on the resonant behavior of our spherical cavity, we show in Fig.8 the computed ratio $E_{\text{inside}}/E_{\text{incident}}$ for $\ell = 10, 20$ and $25$. It is observed that, with an increasing mode number $\ell$, the lowest accessible resonance moves to higher frequencies, and that the $Q$-factor associated with individual resonance lines tends to rise.

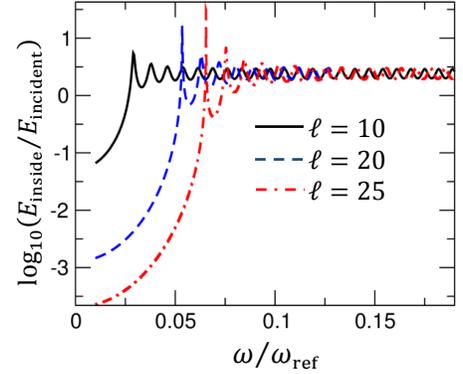

**Fig. 8**. Excitation frequency dependence of the ratio of the $E$-field inside a glass sphere to the incident $E$-field for $\ell = 10$ (solid black), $\ell = 20$ (dashed blue), and $\ell = 25$ (dash-dotted red) TE spherical harmonics.

Finally, Fig.9 shows computed $Q$-factors ($Q = |\omega_q'/\omega_q''|$) for a spherical cavity having $R = 77.5$ μm, $\mu = 1.0$, $n = n' + in''$, and $\ell = 340$. Setting $n' = 1.5$ allows a comparison between the results depicted in Fig.6, where $n'' = 0$, and those in Fig.9, which correspond to $n'' = 10^{-8}$ (blue squares), $10^{-7}$ (red circles), and $10^{-6}$ (black diamonds). These positive values of $n''$ account for the presence of small amounts of absorption within the dielectric sphere. Compared to the case of $n'' = 0$, the resonance frequencies in Fig.9 have not changed by much, but the $Q$-factors of the various resonances are seen to have declined substantially. As expected, the greatest drop in the $Q$-factor is associated with the largest value of $n''$. This is a practically interesting finding, especially in light of the previous result on the $Q$-factors of an idealized cavity, which could reach exceedingly high values. Here we see that accounting for realistic values of optical loss brings down the computed $Q$-factor to the levels observed in experiments [1,2]. This also indicates that the limiting factor in the best spherical resonators available today is most likely the medium properties rather than roughness and other cavity imperfections as one might reasonably presume. Considering that the measured absorption coefficients (e.g., $n'' \cong 10^{-7}$ for a fused silica micro-sphere in the visible optical range) are comparable to the theoretical values of $n''$ needed to bring the $Q$-factor of a perfectly spherical dielectric resonator to within the range of the highest $Q$-factors that are currently accessible to experiments, it is reasonable to conclude that the $Q$-factor-limiting physical effect is in fact absorption within the micro-sphere. Needless to say, scattering from surface roughness and also deleterious effects of inclusions, impurities, and material inhomogeneities could result in mode-mixing, which causes further reduction of the $Q$-factor. Nevertheless, the purity and the polish quality of existing dielectric micro-spheres are such that their observed $Q$-factors indeed appear to be limited by the absorption coefficient $n''$ of the host material.

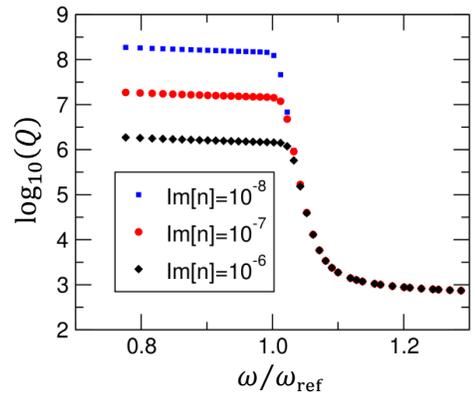

**Fig. 9**. Similar to Fig. 6, except that the refractive index $n = n' + in''$ of the dielectric sphere is now allowed to have a small nonzero imaginary part, $n''$, representing absorption within the material.

**5. Concluding remarks**. Leaky modes contain a wealth of information about the resonant behavior of dielectric cavities, including the lifetimes associated with the light trapped inside the cavity immediately after the source of excitation is turned off. Listed below is a summary of the main results of the present paper, with emphasis placed not only on mathematical aspects but also on the physical attributes of our findings.

1. A dielectric or metallic sphere, when illuminated from the outside or excited internally, contains EM fields. Once the excitation is terminated, the trapped fields inside the sphere decay by leaking out and/or by being absorbed within the sphere. We have identified the complete set of leaky modes, and shown the conditions under which a trapped field can be expressed as a superposition of these leaky modes.

2. We have proven the completeness of these leaky modes under special circumstances, although completeness under more general conditions remains to be demonstrated. We have modelled the dielectric function $\varepsilon(\omega) = n^2(\omega)$ of the spherical particle as a single Lorentz oscillator, thereby treating dispersion and absorption of the material medium in a simple yet physically realistic way. While we have assumed that the sphere is surrounded by free space, the results can be readily extended to the case of a surrounding dielectric medium.

3. Our completeness proof rigorously accounts for realistic dispersion effects, including absorption losses, the existence of branch-cuts associated with the Lorentz oscillator model of the refractive index, and the fact that infinitely many complex poles accumulate in the vicinity of the singular point(s) of the refractive index.

4. We did not invoke the Green function (or tensor) that has been traditionally used to analyze this type of problem. Instead, we relied on the exact solutions of Maxwell's equations to identify the leaky modes, then constructed the modal expansion of an initial field distribution using a straightforward application of the Cauchy theorem of complex analysis; see, e.g., Eq.(19). The explicit formulas derived here for the expansion coefficients allow easy evaluation of the relative contributions to an arbitrary initial distribution (inside the spherical particle) of the various leaky modes; see Eq.(23).

5. With regard to the conventional Green's function approach, we note that one can certainly rewrite Maxwell's equations into an integral equation, and the boundary conditions at infinity are carried by the choice of the Green function. Usually the waves at infinity are either outgoing or incoming, since, for these boundary conditions, Green's functions are easy to find. However, for the specific goal of obtaining leaky-mode expansions of the fields, one also needs to find the leaky modes, then express Green's functions as sums of the leaky modes. This can certainly be done at a formal level, but there are two problems that have to be faced. The first is to find an effective way to calculate the coefficients of the expansion; this, in general, is not a simple problem, and numerical methods might have to be deployed. The second is that the question of convergence of the resulting expansion must be treated separately, as there is nothing in the Green function formulation that would guarantee the convergence of the leaky-mode expansion. To the best of our knowledge, the arguments given in favor of the convergence in the literature do not constitute a rigorous proof, if only because the question of accumulation points of the resonance poles has so far not been analyzed within the Green function framework.

6. As a matter of fact, most works utilizing Green's functions seem to target primarily applications to cavity perturbations rather than address questions of convergence [8,22-27]. In contrast, the approach taken in the present paper is to i) find the leaky modes, ii) find the explicit expansion coefficients of the functions of interest with respect to the modes, and iii) decide on the convergence of the series. Solving Maxwell's equations using scattering boundary conditions in conjunction with Cauchy's theorem addresses the three aforementioned goals in a well-designed, easy-to-use package. Ours is a highly flexible approach in which the design of the leaky modes and the corresponding expansion coefficients are guided by the question of convergence. In fact, rather than being some late-comer to the game, in our approach convergence is actually a design tool.

7. Our numerical results have intimated a close association between resonant behavior and the leaky eigen modes of dielectric spheres. The fact that spherical harmonics with large $\ell$ values are associated with high-$Q$ resonances hints at the importance of electromagnetic angular momentum in relation to the long lifetimes of the modes trapped inside these cavities. In other words, there appears to be a connection between the strength of the circular motion of EM energy inside a cavity and the time it takes for this energy to leak out. We have seen a similar relation between the azimuthal mode number $m$ and the cavity $Q$-factor in the case of cylindrical cavities [15]. In fact, when the radius $R$ and the refractive index $n$ of a dielectric cylinder are the same as those of a sphere, and when the mode number $m$ for the cylinder is the same as the mode number $\ell$ for the sphere ($m = \ell \gg 1$), the plots of $Q$-factor versus resonance frequency for the two cavities are found to be nearly identical.

8. Leaky modes are often characterized as "unphysical" because they seem to carry infinite energy. We have emphasized that the EM field distribution outside the sphere grows exponentially with radial distance, while decaying exponentially with time. The exponential growth with distance, however, is not unphysical, because the fields only extend to a distance $r = ct$ from the sphere's surface, where $t$ is the time elapsed since the external/internal excitation of the spherical particle was terminated. Considering that the leaky modes exist only after the termination of the excitation, the outer tails of the leaky modes within the surrounding medium do not extend to infinity and, therefore, the well-known exponential growth of the field amplitude with distance does not constitute a violation of the law of conservation of energy. (Note that the situation discussed here is completely analogous to that in quantum mechanics; see, e.g., [28].)

9. In Fig.2, we presented a typical map of the leaky frequencies $\omega_q$ in the complex $\omega$-plane, and drew attention to the singular points of this map, which are located at the pole(s) and zero(s) of the refractive index $n(\omega)$ of the spherical particle. It must be emphasized that, when the excited field has a frequency close to the pole(s) of the refractive index, there will be a large number of closely-spaced leaky frequencies that must be included in any physically meaningful expansion of the initial field distribution.

10. We have provided several numerical examples, some with artificial parameter values to emphasize the mathematical aspects of the leaky mode expansion (e.g., Figs. 2-5), and some with physically realistic parameter values (e.g., Figs. 6-9) in order to draw attention to the behavior of leaky modes in problems of practical interest.

11. Finally, it is interesting to note that small amounts of absorption or loss can dramatically suppress the $Q$-factors of a solid dielectric sphere at large $\ell$ and in the vicinity of the cutoff frequency, as revealed by a comparison between Fig.6 and Fig.9. This finding indicates that the $Q$-factors occurring in practice might be actually limited by the material properties rather than the particle's surface quality.

In conclusion, the present paper has described a general approach to analyzing and computing the leaky modes of solid dielectric spheres. In Sec.3, we presented the outlines of a completeness proof for expanding certain initial field distributions as a sum over leaky modes. Mathematical details and some of the subtleties associated with the series convergence were either skipped over or mentioned only briefly. These subtleties, which revolve around the behavior of the accumulated poles of the function $G(\omega)$ of Eq.(18) when $\omega$ approaches the poles $\Omega_{3,4}$ of the refractive index $n(\omega)$, and also when $\omega \to \infty$, will be the subject of a forthcoming mathematics-oriented paper.

**Acknowledgement**. This work has been supported in part by the AFOSR grant No. FA9550-13-1-0228.

## Appendix

We show that $F(\omega)$ of Eq.(17) approaches a constant when $\omega \to 0$. In the limit $z \to 0$, we have

$$J_\nu(z) \to \frac{(z/2)^\nu}{\Gamma(\nu+1)}. \tag{A1}$$

$$Y_\nu(z) \to \frac{(z/2)^\nu}{\tan(\nu\pi)\Gamma(1+\nu)} - \frac{(z/2)^{-\nu}}{\sin(\nu\pi)\Gamma(1-\nu)}; \quad (\nu \neq \text{an integer}). \tag{A2}$$

Therefore, when $\omega \to 0$, considering that $k_0 = \omega/c \to 0$, we will have

$$F(\omega) = nk_0 R \mathcal{H}^{(1)}_{\ell+\frac{1}{2}}(k_0 R) J_{\ell+3/2}(nk_0 R) + [(\mu-1)(\ell+1)\mathcal{H}^{(1)}_{\ell+\frac{1}{2}}(k_0 R) - \mu k_0 R \mathcal{H}^{(1)}_{\ell+3/2}(k_0 R)] J_{\ell+\frac{1}{2}}(nk_0 R)$$

$$= nk_0 R J_{\ell+\frac{1}{2}}(k_0 R) J_{\ell+3/2}(nk_0 R) + [(\mu-1)(\ell+1) J_{\ell+\frac{1}{2}}(k_0 R) - \mu k_0 R J_{\ell+3/2}(k_0 R)] J_{\ell+\frac{1}{2}}(nk_0 R)$$

$$+ ink_0 R Y_{\ell+\frac{1}{2}}(k_0 R) J_{\ell+3/2}(nk_0 R) + i[(\mu-1)(\ell+1) Y_{\ell+\frac{1}{2}}(k_0 R) - \mu k_0 R Y_{\ell+3/2}(k_0 R)] J_{\ell+\frac{1}{2}}(nk_0 R)$$

$$\to \frac{i(-1)^{\ell+1} nk_0 R}{\Gamma(\frac{1}{2}-\ell)\Gamma(\ell+5/2)} (\tfrac{1}{2}k_0 R)^{-(\ell+\frac{1}{2})} (\tfrac{1}{2}nk_0 R)^{\ell+3/2}$$

$$+ \frac{i}{\Gamma(\ell+3/2)} \left[ \frac{(-1)^{\ell+1}(\mu-1)(\ell+1)}{\Gamma(\frac{1}{2}-\ell)} (\tfrac{1}{2}k_0 R)^{-(\ell+\frac{1}{2})} - \frac{(-1)^\ell \mu k_0 R}{\Gamma(-\frac{1}{2}-\ell)} (\tfrac{1}{2}k_0 R)^{-(\ell+3/2)} \right] (\tfrac{1}{2}nk_0 R)^{\ell+\frac{1}{2}}$$

$$\to \frac{i(-1)^{\ell+1} n^{\ell+\frac{1}{2}}}{\Gamma(\ell+3/2)} \left[ \frac{(\mu-1)(\ell+1)}{\Gamma(\frac{1}{2}-\ell)} + \frac{2\mu}{\Gamma(-\frac{1}{2}-\ell)} \right] = \frac{i(-1)^\ell [1+\ell+\ell\mu(0)][n(0)]^{\ell+\frac{1}{2}}}{(\ell+\frac{1}{2})\Gamma(\frac{1}{2}+\ell)\Gamma(\frac{1}{2}-\ell)}. \tag{A3}$$

The identity $\Gamma(x+1) = x\Gamma(x)$ has been used in the above derivation. We may now invoke the identity $\Gamma(\frac{1}{2}+x)\Gamma(\frac{1}{2}-x) = \pi/\cos(\pi x)$ to arrive at

$$\lim_{\omega \to 0} F(\omega) = i\left[\sqrt{\mu(0)\varepsilon(0)}\right]^{\ell+\frac{1}{2}} [1 + \ell + \ell\mu(0)]/[(\ell+\frac{1}{2})\pi]. \tag{A4}$$

It is seen that $F(\omega)$ has no poles at $\omega = 0$, which indicates that, in the vicinity of $\omega = 0$, the function $G(\omega)$ is not singular.